\documentclass[preprintnumbers,amsmath,amssymb,showpacs]{revtex4}
\usepackage{epsfig}
\usepackage{color}
\usepackage[colorlinks,urlcolor=blue,citecolor=blue]{hyperref}
\begin{document}
\setlength{\voffset}{1.0cm}
\title{Non-Abelian twisted kinks in chiral Gross-Neveu model with isospin}
\author{Michael Thies\footnote{michael.thies@gravity.fau.de}}
\affiliation{Institut f\"ur  Theoretische Physik, Universit\"at Erlangen-N\"urnberg, D-91058, Erlangen, Germany}
\date{\today}
\begin{abstract}
The two-dimensional, massless Gross-Neveu model with $N_c$ colors and SU(2) isospin is studied analytically in the large $N_c$ limit.
The chiral SU(2)$_L \times$SU(2)$_R$ symmetry is broken spontaneously in the vacuum.
Twisted kinks connecting two arbitrary points on the vacuum manifold $S^3$ are constructed, and their properties are explored.
The phase diagram as a function of temperature, baryon- and isospin chemical potential is discussed, with special emphasis on
inhomogeneous phases. The preferred form of the condensate is a product of the real kink crystal and the chiral spiral.
Kink-kink scattering is solved, using the general solution of the multicomponent Bogoliubov-de Gennes
equation recently presented by Takahashi.
\end{abstract}
\maketitle
\section{Introduction}
\label{sect1}
Gross-Neveu (GN) models \cite{L1} with massless fermions come with either discrete or continuous chiral symmetry.
The most widely studied version with continuous chiral symmetry has the Lagrangian
\begin{equation}
{\cal L}_{\rm NJL} =   \bar{\psi} i\partial \!\!\!/ \psi + \frac{g^2}{2} \left[ \left( \bar{\psi} \psi \right)^2 + \left( \bar{\psi} i \gamma_5\psi \right)^2\right]  ,
\label{A1}
\end{equation}
where summation over $N$ fermion species is implied. It is commonly referred to as ``chiral GN model" or else as ``two-dimensional Nambu--Jona Lasinio model"
(NJL$_2$). The Lagrangian (\ref{A1}) is invariant under the Abelian  chiral symmetry group U(1)$_L \times$U(1)$_R$. In the real world, low energy strong interactions are 
governed by an approximate SU(2)$_L\times$SU(2)$_R$ chiral symmetry, reflecting the presence of two light quark flavors. Spontaneous symmetry breaking
then gives rise to an isotriplet of light pions. As the original NJL model in 3+1 dimensions \cite{L2} was designed as a phenomenological
model of strong interactions, it has also been endowed with the non-Abelian SU(2)$_L \times$SU(2)$_R$ chiral group. The Lagrangian of the 
NJL model with isospin (isoNJL) differs from expression (\ref{A1}) in the pseudoscalar-pseudoscalar interaction term and reads   
\begin{equation}
{\cal L}_{\rm isoNJL} =   \bar{\psi} i\partial \!\!\!/ \psi + \frac{g^2}{2} \left[\left(  \bar{\psi} \psi \right)^2 +
\left(  \bar{\psi} i \gamma_5 \vec{\tau} \psi \right)^2\right] \, .
\label{A2}
\end{equation}
The fermion fields $\psi$ carry ``color" (1...$N_c$), ``isospin" (1...2) and Dirac (1...2) indices, which will all be suppressed whenever possible.
Since the isoNJL model is not a gauge theory, color and isospin should both be regarded as ``flavors". Isospin is distinguished from the other flavors by the fact that
it enters the chiral symmetry group, by choice of the four-fermion interaction term. 

As long as one is interested in GN models as purely theoretical laboratories, there is nothing wrong in simplifying the chiral group to U(1)$_L \times$U(1)$_R$
in 1+1 dimensions. However, in recent years, the barrier between toy models and more phenomenological approaches in 3+1 dimensions has become more permeable. 
As a matter of fact, several results from 1+1 dimensional model studies have played a role in strong interaction physics, notably in
questions of hot and dense matter. We only mention the ``quarkyonic phase" of dense matter due to McLerran, Pisarski and collaborators \cite{L3,L4,L5}, 
akin to the ``chiral spiral" type soliton 
crystal of the NJL$_2$ model \cite{L6},
or Nickels finding \cite{L7,L8,L9} that the soliton crystal of the 2d GN model with discrete chiral symmetry \cite{L10,L11} is a good variational ansatz for the 3d isoNJL model with
continuous chiral symmetry. Apart from general theoretical interest, this motivates us to study the isoNJL$_2$ model with non-Abelian chiral symmetry in
1+1 dimensions more systematically in the present paper.  

Whereas the standard NJL$_2$ model is by now very well understood, both as far as thermodynamics \cite{L12,L13} and soliton dynamics \cite{L14,L15,L16,L17} are concerned,
investigations of the isoNJL$_2$ model in the existing literature deal mostly with its phase diagram and are less complete. Ebert, Klimenko and collaborators 
have first studied homogeneous phases with baryon and isospin chemical potentials \cite{L18} and found a rather complex phase structure.
Subsequently, inhomogeneous condensates were taken into account as well, either in the scalar-neutral pion \cite{L19} or the charged pion \cite{L20} sector.
In spite of various insights and interesting partial results, a definite picture of  
the full phase diagram with two chemical potentials did not yet emerge.
  
In a recent unbiased numerical study, Heinz et al. \cite{L21}  found the puzzling result that the phase diagram of the isoNJL$_2$ model as
a function of temperature and baryon chemical potential coincides with that of the GN model with discrete chiral symmetry \cite{L22,L23}.
Needless to say, this observation calls for an analytic understanding. Aside from these results on the phase diagram, little if anything
seems to be known about multifermion bound states and the dynamics of possible solitons in the isoNJL$_2$ model,
in strong contrast to the standard NJL$_2$ model.

From our previous experience with GN-type models, we believe that the key to both hadron structure and hot and dense matter lies in
``twisted kinks", the most elementary hadrons whose condensates interpolate between two different vacua as a function of $x$.
The original twisted kink of Shei \cite{L24} connects two points on the chiral circle ($S^1$), characteristic for the U(1)$_L \times$U(1)$_R$ chiral symmetry
of the NJL$_2$ model.
It can be viewed as the elementary building block of all hadrons in GN models with either discrete or continuous chiral symmetry.
What is the corresponding object in the SU(2) chirally symmetric isoNJL model,
where the vacuum manifold is not $S^1$, but $S^3$? This is the main question we propose to address here. Such a study is expected to be 
technically more involved, as the time dependent Hartree-Fock (TDHF) equation needs to be generalized to two (isospin-)components. 
Fortunately, in condensed matter physics, the corresponding problem for the Bogoliubov-de Gennes (BdG) equation for 1d superconductors has
recently been solved in great generality by Takahashi \cite{L25}. His paper has turned out to be very useful for the part of our present study dealing with
the scattering of twisted kinks.

The plan of our paper is as follows. In Sect.~\ref{sect2}, we outline the formalism with a focus on the symmetries and the TDHF approach.
In Sect.~\ref{sect3}, we determine the vacua and recall the gap equation of GN-type models. In Sect.~\ref{sect4} we construct the non-Abelian twisted kink
of the isoNJL$_2$ model and relate it to the original Abelian kink of Shei. 
Sect.~\ref{sect5} addresses the question: What is the analogue of the chiral spiral in the model with isospin? Then we put together everything which can
be said about the phase diagram of the isoNJL$_2$ model.
Sect.~\ref{sect6} contains the solution of the kink-kink scattering problem, based on the recent work of Takahashi on the multi-component BdG equation
in condensed matter physics \cite{L25}. 
We finish with a brief summary and our conclusions in Sect.~\ref{sect7}. Appendix 1 summarizes well-known results for the kink and kink-kink scattering
within the standard NJL$_2$ model to which we frequently refer in the present article. In Appendix 2 we gather some technical details of how to apply
the formalism of Ref.~\cite{L25} to the present problem.

\section{General formalism}
\label{sect2}
Consider the Lagrangian (\ref{A2}) of the isoNJL$_2$ model. Apart from the SU($N_c$) and Poincar\'e symmetries, the Lagrangian is invariant under 
U$_V(1)$ (conservation of fermion number)
\begin{equation}
\psi \to e^{i\alpha} \psi, \quad
\partial_{\mu} j^{\mu} = \partial_{\mu} \bar{\psi} \gamma^{\mu} \psi = 0 ,
\label{B1}
\end{equation}
and chiral SU(2)$_L \times $SU(2)$_R$ symmetry with conserved vector/isovector and axial vector/isovector currents,
\begin{eqnarray}
\psi_L & \to &  e^{i \vec{\alpha}\vec{\tau}} \psi_L,  \quad \psi_R \to e^{i \vec{\beta}\vec{\tau}} \psi_R,
\nonumber \\
\partial_{\mu} j_a^{\mu} & = & \partial_{\mu} \bar{\psi} \gamma^{\mu}\frac{\tau_a}{2} \psi = 0,
\nonumber \\
\partial_{\mu} j_{5,a}^{\mu} & = & \partial_{\mu} \bar{\psi} \gamma^{\mu}\gamma_5 \frac{\tau_a}{2} \psi = 0.
\label{B2}
\end{eqnarray}
The isospin operators $\tau_a$ have the same form as the Pauli matrices $\sigma_a$.
The isoscalar axial current $\bar{\psi} \gamma^{\mu} \gamma_5 \psi$ is not conserved, unlike in the standard chiral GN model, Eq.~(\ref{A1}).
Specific for 1+1 dimensions is the fact that vector and axial vector currents are not independent, but related as follows,
\begin{equation}
j_5^0 = j^1, \quad j_5^1=j^0.
\label{B3}
\end{equation}
In the standard NJL$_2$ model where both $j$ and $j_5$ are conserved, this can be used to justify the relations
\begin{equation}
j^{\mu} = \epsilon^{\mu \nu} \partial_{\nu} \phi, \quad j_5^{\mu} = \partial^{\mu} \phi,
\label{B4}
\end{equation}
and to interpret the conservation of the axial current as Klein-Gordon equation of a massless, pseudoscalar boson,
\begin{equation}
\partial_{\mu} j_5^{\mu} = \partial_{\mu} \partial^{\mu} \phi =0.
\label{B5}
\end{equation}
These various relations can also be used to understand why a static fermion charge density cannot be localized in the standard NJL$_2$ model, but has to be
spatially constant \cite{L26}. In the isoNJL$_2$ model, similar considerations hold for the isovector currents,
\begin{eqnarray}
j_{5,a}^{0} & = &  j_a^{1}, \quad j_{5,a}^{1}=j_a^{0},
\nonumber \\
j_a^{\mu} &  = & \epsilon^{\mu \nu}\partial_{\nu} \pi_a, \quad j_{5,a}^{\mu} = \partial^{\mu} \pi_a.
\label{B6}
\end{eqnarray} 
Here, axial vector/isovector  current conservation is equivalent to the Klein-Gordon equation of a triplet of massless 
(pseudoscalar, isovector) ``pions",
\begin{equation}
\partial_{\mu} j_{5,a}^{\mu} = \partial_{\mu} \partial^{\mu} \pi_a =0.
\label{B7}
\end{equation}
Furthermore, the isovector charge density $j_a^{0}$ of time-independent states is delocalized,
whereas there is no obstruction to localizing baryonic charge. 

Here as in most previous studies of GN models we are interested in the 't~Hooft limit ($N_c\to \infty, N_c g^2 = {\rm const.}$).
This amounts to writing down the Euler-Lagrange equations for the fermion fields, replacing color singlet bilinears by
$c$-number condensates (TDHF equation)
\begin{equation}
\left( i \gamma^{\mu} \partial_{\mu} - S - i \gamma_5 P_a \tau_a \right) \psi = 0
\label{B8}
\end{equation}
with
\begin{equation}
S = - g^2 \langle \bar{\psi} \psi \rangle, \quad P_a = - g^2 \langle \bar{\psi} i \gamma_5 \tau_a \psi \rangle. 
\label{B9}
\end{equation}
$S$ can be thought of as the $c$-number part of a scalar field $\sigma$, $P_a$ as the $c$-number part of a pseudoscalar/isovector field $\pi_a$. 
We use the following chiral representation of Dirac matrices,
\begin{equation}
\gamma^0 = \sigma_1, \quad \gamma^1 = i \sigma_2, \quad \gamma_5 = \gamma^0 \gamma^1 = - \sigma_3.
\label{B10}
\end{equation}
The Dirac-TDHF equation in Hamiltonian form then becomes
\begin{equation}
i \partial_t \left( \begin{array}{c} \psi_{1,1} \\ \psi_{1,2} \\ \psi_{2,1} \\ \psi_{2,2} \end{array} \right)
= \left( \begin{array}{cccc} i \partial_x & 0 & {\cal D}^*  & {\cal C}^* \\ 0 & i \partial_x & - {\cal C} &  {\cal D} \\
{\cal D} & - {\cal C}^* & -i \partial_x & 0 \\ {\cal C}  & {\cal D}^*  & 0 & -i \partial_x \end{array} \right)
\left( \begin{array}{c} \psi_{1,1} \\ \psi_{1,2} \\ \psi_{2,1} \\ \psi_{2,2} \end{array} \right).
\label{B11}
\end{equation}
The first subscript on the spinor components is the Dirac index ($1=L,2=R$), the 2nd one the isospin index
(1 = up, 2 = down), and we have introduced two complex condensates
\begin{eqnarray}
{\cal D}  & = & S-iP_3,
\nonumber \\
{\cal C} & = & P_2-iP_1,
\label{B12}
\end{eqnarray}
related to the scalar meson and neutral pion ($\cal D$) and charged pions ($\cal C$), respectively.
Introducing light cone coordinates
\begin{equation}
z=x-t, \quad \bar{z} =x+t, \quad \partial_0 = \bar{\partial}-\partial, \quad \partial_1 = \bar{\partial}+\partial,
\label{B13}
\end{equation}
Eq.~(\ref{B11}) can be cast into  the covariant form of a two-component Dirac-TDHF (or BdG) equation,
\begin{eqnarray}
2i \bar{\partial} \Psi_2 & = &  \Delta \Psi_1,
\nonumber \\
2 i \partial \Psi_1 & = & - \Delta^{\dagger} \Psi_2,
\label{B14}
\end{eqnarray}
with the notation in isospin-space
\begin{equation}
\Psi_i = \left( \begin{array}{c} \psi_{i,1} \\ \psi_{i,2} \end{array} \right), \quad \Delta = \left( \begin{array}{rr} {\cal D} & -{\cal C}^* \\
{\cal C} &  {\cal D}^* \end{array} \right).
\label{B15}
\end{equation}
The self-consistency conditions following from (\ref{B9}) are
\begin{eqnarray}
{\cal D} & = &  - 2 N_cg^2 \sum^{\rm occ} \left( \psi_{1,1}^* \psi_{2,1}+ \psi_{2,2}^* \psi_{1,2} \right),
\nonumber \\
{\cal C}& = & -2 N_cg^2 \sum^{\rm occ} \left( \psi_{1,1}^* \psi_{2,2} -  \psi_{2,1}^* \psi_{1,2} \right),
\label{B16}
\end{eqnarray}
where the summation is over all occupied states, including the Dirac sea.
It is instructive to cast Eq.~(\ref{B16})  into the following matrix form
\begin{equation}
\Delta = - 2 N_cg^2 \sum^{\rm occ} \left[ \left( \begin{array}{c} \psi_{2,1} \\ \psi_{2,2} \end{array} \right) \left( \psi_{1,1}^*, \psi_{1,2}^* \right) 
+  \left( \begin{array}{c} \psi_{2,2}^* \\ -\psi_{2,1}^* \end{array} \right) \left( \psi_{1,2}, - \psi_{1,1} \right) \right]
\label{B17}
\end{equation}
or, in terms of the isospinors introduced in Eq.~(\ref{B15}), 
\begin{equation}
\Delta = - 2 N_c g^2 \sum^{\rm occ} \left( \Psi_2 \Psi_1^{\dagger} + \tilde{\Psi}_2 \tilde{\Psi}_1^{\dagger}\right), \quad \tilde{\Psi}_k = i \tau_2 \Psi_k^* .
\label{B18}
\end{equation}
Due to the well-known fact that
\begin{equation}
\tau_2 U \tau_2 = U^*
\label{B19}
\end{equation}
for any SU(2) matrix $U$ in the fundamental representation, $\Psi_k$ and $\tilde{\Psi}_k$ transform identically under
SU(2)$_L \times$SU(2)$_R$ chiral transformations. Thus self-consistency is manifestly preserved under global
chiral transformations
\begin{equation}
\Psi_1\to U_L \Psi_1, \quad \Psi_2 \to U_R \Psi_2, \quad \Delta \to U_R \Delta U_L^{\dagger}.
\label{B20}
\end{equation}
This will be exploited below to simplify several computations.

\section{Vacua and renormalization}
\label{sect3}
For homogeneous condensates ($S,P_a$), the solution of the Dirac-HF equation is trivial with fermion spectrum
\begin{equation}
\omega = \pm \sqrt{m^2+k^2} .
\label{C1}
\end{equation}
Each state is twofold degenerate due to isospin, and the physical fermion mass satisfies
\begin{equation}
m^2 = S^2 + P_a P_a.
\label{C2}
\end{equation}
We choose units such that $m=1$ from here on. A non-vanishing condensate signals spontaneous symmetry breaking of chiral
symmetry, only possible in the large $N_c$ limit in 2 dimensions. The vacuum manifold is the group SU(2) or, equivalently, $S^3$,
as evidenced by the fact that
\begin{equation}
\Delta = \left( \begin{array}{rr} S-i P_3 & -P_2-i P_1 \\ P_2 - i P_1 & S+iP_3 \end{array} \right) \in {\rm SU(2)\ for\ } S^2+P_a P_a = 1.
\label{C3}
\end{equation}
By a global chiral transformation, $\Delta \in$ SU(2) can always be 
``rotated" into the unit matrix.
The Dirac equation (\ref{B14}) then simplifies to 
\begin{eqnarray}
2i \bar{\partial} \Psi_2 & = &   \Psi_1,
\nonumber \\
2 i \partial \Psi_1 & = & -  \Psi_2.
\label{C4}
\end{eqnarray}
This reduces the vacuum problem to that of the standard GN model, except for a degeneracy factor of 2 due to isospin.
Renormalization is done in the familiar way with $2N_c$ playing the role of the number of flavors $N$ in $Ng^2$,
\begin{equation}
1 + \frac{Ng^2}{\pi} \ln \frac{m}{\Lambda} = 0, \quad (N=2 N_c).
\label{C5}
\end{equation}
For the derivation of the gap equation (\ref{C5}) with momentum cutoff $\Lambda/2$, see e.g. Ref. \cite{L12}. The gap equation 
yields the relation between the (dimensionless) bare coupling
constant, the momentum cutoff
and the physical fermion mass, and is characteristic for a theory without a scale like the massless GN model (dimensional transmutation).

\section{Non-Abelian twisted kink}
\label{sect4}
In the standard NJL$_2$ model, twisted kinks are the key to understanding soliton bound states, breathers and
scattering problems. What is the analogue object in the isoNJL$_2$ model? This is the topic of the present section.

The most general twisted kink of the isoNJL$_2$ model should interpolate between two arbitrary points on the vacuum manifold $S^3$. Let us parametrize these 
points as
\begin{eqnarray}
\lim_{x \to - \infty} \Delta & = &  \Delta_- = e^{-2i \vec{\phi}_-\vec{\tau}},
\nonumber \\
\lim_{x \to +  \infty} \Delta & = &  \Delta_+ = e^{-2i \vec{\phi}_+\vec{\tau}} e^{-2i \vec{\phi}_-\vec{\tau}}.
\label{D1}
\end{eqnarray}
The $-$ sign and the factor of 2 in the exponents have been inserted to match the convention used in Ref. \cite{L17} for the Abelian twisted kink. 
Parametrizing $\Delta_+$ as a single SU(2) factor might seem more natural, but this would lead to algebraic
complications later on due to the composition law for finite SU(2) transformations.  
Next we simplify $\Delta_{\pm}$ as
much as possible by means of a global chiral transformation,
\begin{equation}
\Delta_{\pm} \to U_R \Delta_{\pm} U_L^{\dagger}, \quad U_{L,R} \in {\rm SU(2)}.
\label{D2}
\end{equation}
The choice
\begin{equation}
U_R = \Omega_+, \quad U_L = \Omega_+  e^{-2i \vec{\phi}_-\vec{\tau}}
\label{D3}
\end{equation}
where $\Omega_+$ is a matrix which diagonalizes $\vec{\phi}_+ \vec{\tau}$,
\begin{equation}
\Omega_+ \vec{\phi}_+ \vec{\tau}\, \Omega_+^{\dagger} = \phi_+ \tau_3,
\label{D4} 
\end{equation}
maps $\Delta_-$ onto the unit matrix while diagonalizing $\Delta_+$,
\begin{equation}
\Delta_- \to 1, \quad \Delta_+ \to  e^{-2i \phi_+ \tau_3}.
\label{D5}
\end{equation}
If we assume that $\Delta$ stays diagonal during the whole trajectory from $\Delta_-$ to $\Delta_+$, the two isospin channels
decouple in the TDHF equation and the problem can be reduced to the known kink of the standard NJL$_2$ model.
This assumption is justified a posteriori by verifying that the solution thus obtained has all the desired properties. 

In the isospin up channel, the potential
${\cal D} = S - iP_3$ plays the role of the scalar-pseudoscalar potential $\Delta= S-iP$ of the 
standard NJL$_2$ model 
\begin{equation}
2i\bar{\partial} \psi_{2,1}  =  {\cal D} \psi_{1,1}, \quad 2i \partial \psi_{1,1} = -{\cal D}^* \psi_{2,1}.
\label{D6}
\end{equation}
For isospin down, ${\cal D}$ gets replaced by ${\cal D}^*$, i.e., we are faced with the charge conjugate problem,
\begin{equation}
2i\bar{\partial} \psi_{2,2}  = {\cal D}^* \psi_{1,2}, \quad 2i \partial \psi_{1,2} = -{\cal D} \psi_{2,2}.
\label{D7}
\end{equation}
Thus the original problem of the non-Abelian twisted kink has been reduced to solving the standard NJL model for the kink and its charge conjugate.
These solutions are well known since the early work \cite{L24}. For convenience, they are summarized in Appendix 1 in the notation of \cite{L17,L18}. 
Nonetheless, the resulting kink of the isoNJL$_2$ model is non-trivial and has novel physical 
properties as compared to the NJL$_2$ twisted kink. 
This is due to the locking of a kink and its charge conjugate in the two isospin channels and, related to this, 
the self-consistency conditions (\ref{B16}) of the isoNJL$_2$ model, different from the one
of the standard NJL$_2$ model.
The potential and spinors can be taken over literally from the NJL$_2$ model, Appendix 1. We denote $\phi_+$
by $\phi$ from now on, since there is only one twist angle due to our choice of internal coordinate frame.
The twisted kink potential then assumes the form
\begin{equation}
\Delta = \frac{1}{1+V} \left( \begin{array}{cc} 1+e^{-2i\phi}V & 0 \\ 0 & 1+e^{2i\phi}V \end{array} \right). 
\label{D8}
\end{equation}
The space-time dependence is carried by the function $V$. For the kink at rest for instance,
\begin{equation}
V= e^{ (z+\bar{z})\sin \phi } = e^{2 x \sin \phi}.
\label{D9}
\end{equation}
Since $\Delta$ is Lorentz scalar, the general form of $V$ as given in Appendix 1, Eq.~(\ref{Q5}), simply amounts to
replacing the spatial coordinate $x$ by the boosted one.

Eq.~(\ref{D8}) is a special case of the most general kink potential, obtained by undoing the chiral transformation (\ref{D2}),
\begin{equation}
\Delta = \frac{\Delta_- + \Delta_+ V}{1+V}, \quad \Delta_{\pm} \in {\rm SU(2)}.
\label{D10}
\end{equation}
The matrix $\Delta$ is in general non-diagonal, so that the two isospin channels mix. $V$ has the same form as before.
In the case that $\Delta_{\pm}$ in (\ref{D10}) are prescribed, the angle $\phi$ entering $V$ can be determined as follows,
\begin{equation}
\phi = \arccos \frac{1}{2} {\rm Tr} \left( \Delta_+ \Delta_-^{\dagger} \right).
\label{D11}
\end{equation}

Let us go back to the simple, diagonal form of $\Delta_{\pm}$ of Eq.~(\ref{D5}). 
The spinors can be constructed from the ones in Appendix 1. 
Here we focus on the new aspects due to the self-consistency condition.
For the isospin up channel, the condensate is given by
\begin{eqnarray}
2 \sum^{\rm occ} \psi_{1,1}^* \psi_{2,1} & = & - \frac{\ln \Lambda}{\pi} \left( \frac{ 1+e^{-2i\phi}V}{1+V} \right)N_c
\nonumber \\
& & - \left( \frac{\phi}{\pi} - \nu_1\right) 2 \sin \phi e^{-i\phi} \frac{V}{(1+V)^2}N_c
\label{D12}
\end{eqnarray}
where $\nu_1$ is the occupation fraction of the bound state. It multiplies the bound state contribution.
The remaining terms are due to the continuum states in the Dirac sea. The fermion density is
\begin{equation}
\sum^{\rm occ} \left( |\psi_{1,1}|^2 + |\psi_{2,1}|^2 \right) = \left(\nu_1-\frac{\phi}{\pi} \right) 2 \sin \phi \frac{V}{(1+V)^2}N_c .
\label{D13}
\end{equation}
For the standard NJL$_2$ model the divergent term in the condensate ($\sim \ln \Lambda$) yields self-consistency if one invokes 
the vacuum gap equation. The finite piece vanishes provided that $\nu_1=\phi/\pi$. The same condition yields
a vanishing fermion density, a direct consequence of the conservation of the axial current.
Now consider the isospin down channel. Charge conjugation replaces $\phi$ with $\pi-\phi$
and switches the sign of the bound state energy (from $\cos \phi$ to $-\cos \phi$). $V$ is unchanged, but relations (\ref{D12},\ref{D13})
go over into
\begin{eqnarray}
2 \sum^{\rm occ} \psi_{1,2}^* \psi_{2,2} & = & - \frac{\ln \Lambda}{\pi} \left( \frac{ 1+e^{2i\phi}V}{1+V}\right)N_c
\nonumber \\
& & + \left( 1-\frac{\phi}{\pi} - \nu_2\right) 2 \sin \phi e^{i\phi} \frac{V}{(1+V)^2}N_c
\label{D14}
\end{eqnarray}
and
\begin{equation}
\sum^{\rm occ} \left( |\psi_{1,2}|^2 + |\psi_{2,2}|^2 \right)  =  \left(\nu_2-1+\frac{\phi}{\pi} \right) 2 \sin \phi \frac{V}{(1+V)^2} N_c. 
\label{D15}
\end{equation}
Here the self-consistency condition in the standard NJL$_2$ model would be $\nu_2 = 1- \phi/\pi$, implying again 
vanishing fermion density.
The self-consistency condition for the isoNJL$_2$ model (\ref{B16}) involves the sum of (\ref{D12}) and the complex conjugate
of (\ref{D14}), with the following result for the condensate
\begin{eqnarray}
2 \sum^{\rm occ} \left( \psi_{1,1}^*\psi_{2,1} + \psi_{2,2}^* \psi_{1,2}\right) & = & - 2 \frac{\ln \Lambda}{\pi}\left( \frac{1+e^{-2i\phi}V}{1+V} \right)N_c
\nonumber \\
& & + \left(  \nu_1-\nu_2 + 1 -\frac{2\phi}{\pi} \right) 2 \sin \phi e^{-i\phi} \frac{V}{(1+V)^2}N_c .
\label{D16}
\end{eqnarray}
The factor of 2 in the divergent part ($\sim \ln \Lambda$) is just what is needed to match the gap equation ($N=2 N_c$) of the isoNJL$_2$ model. Self-consistency
then demands that the finite term vanishes. This reduces to the difference of the self-consistency conditions for the standard NJL$_2$  model and its 
charge conjugate,
\begin{equation}
\nu_1-\nu_2 +1 - \frac{2\phi}{\pi} = 0. 
\label{D17}
\end{equation}
The total fermion density, on the other hand, is the sum of (\ref{D13}) and (\ref{D15}),
\begin{equation}
\sum^{\rm occ} \left( |\psi_{1,1}|^2+|\psi_{2,1}|^2+|\psi_{1,2}|^2+|\psi_{2,2}|^2\right) = \left( \nu_1+\nu_2-1 \right) 2 \sin \phi \frac{V}{(1+V)^2}N_c ,
\label{D18}
\end{equation}
with the corresponding total fermion number 
\begin{equation}
N_f = (\nu_1+\nu_2-1)N_c.
\label{D19}
\end{equation}
As expected on the basis of the symmetries discussed in Sect.~\ref{sect2}, the fermion density of the twisted kink is non-zero in the isoNJL$_2$ model.
We anticipate however that the isospin density $\rho_a = \langle \psi^{\dagger} \tau_a \psi \rangle $ vanishes identically. This is trivial for the 1- and 2-components
 in the present case, since each spinor has only one non-vanishing isospin component. For the 3-component, it can 
be checked by an explicit computation,
\begin{equation}
\sum^{\rm occ} \left( |\psi_{1,1}|^2+|\psi_{2,1}|^2-|\psi_{1,2}|^2-|\psi_{2,2}|^2\right) = \left( \nu_1-\nu_2+1- \frac{2\phi}{\pi} \right) 2 \sin \phi \frac{V}{(1+V)^2} N_c .
\label{D20}
\end{equation}
This vanishes indeed on account of the self-consistency condition (\ref{D17}).  

Trivially, the mass of the kink,  is twice the mass of the kink in the
standard NJL$_2$ model for $N_c$ flavors,
\begin{equation}
M=\frac{2N_c\sin \phi}{\pi} = \frac{N \sin\phi}{\pi}.
\label{D21}
\end{equation}
This holds independently of the direction of the rotation axes. Only the ``intrinsic" twist angle $\phi$ as given in Eq.~(\ref{D11}) enters.
Hence the kink mass per flavor is the same as in the standard NJL$_2$ model.

The maximum fermion number carried by the kink, $N_f=N_c$, corresponds to $\phi=\pi/2$. In this case, each isospin component reduces to the real
kink of the Gross-Neveu model. Since ${\cal D}$ is real, the matrix $\Delta$ is the unit matrix times 
the GN kink potential $S(x)$. Any global chiral transformations replaces the unit matrix by a constant
SU(2) matrix, keeping the same factor $S(x)$ in all 4 components. 
All of these solutions are strictly equivalent, as are the different points on the $S^3$ vacuum manifold.
This type of kink is important for understanding the structure of dense matter in the isoNJL$_2$ model in the following section,
as it can be regarded as the low density limit of baryonic matter.

\section{Massless hadrons, chiral spiral, and phase diagram}
\label{sect5}
A remarkable property of the standard NJL$_2$ model is the existence of massless, delocalized  baryons, which are also the seed of the ``chiral spiral"
type soliton crystal in hot and dense matter \cite{L6}. Some authors have already searched for similar structures in the isoNJL model \cite{L19,L20}. What can
be said about this in the light of the foregoing discussion?

The basic mechanism how to generate and understand the chiral spiral in the standard NJL$_2$ model may be described as follows: 
Start from the time independent Hartree-Fock (HF) equation for homogeneous condensates $S=1,P=0$ appropriate for the vacuum
and perform a linearly $x$-dependent chiral transformation
\begin{equation}
\psi \to \psi' = e^{ibx\gamma_5}\psi.
\label{E1}
\end{equation}
This changes the constant potential into the chiral
spiral potential $S-iP = e^{2ibx}$
and shifts the single particle spectrum rigidly by the amount $b$.  If one evaluates the energy and fermion density of such a state
(relative to the vacuum) with a careful regularization of the Dirac sea, one finds 
\begin{equation}
{\cal E} = \frac{b^2}{2\pi} N_c , \quad \rho = \frac{b}{\pi} N_c. 
\label{E2}
\end{equation}  
This configuration is the energetically most favorable state with finite baryon density at $T=0$. Finite temperature and chemical potential can be dealt with in a similar way
by transforming away the chemical potential by such a local chiral transformation, mapping the point ($T, \mu$) of the phase diagram onto the point ($T,0$) \cite{L12}.
The massless baryon can be thought of as the limit where there is just one whole turn of the chiral spiral along the $x$-axis.
The baryon density is constant in space, in accordance with the conservation of axial current.

Let us start to try to adapt this procedure to the isoNJL$_2$ model. The original transformation of the chiral spiral case, (\ref{E1}),  is no longer allowed, as it
would induce an isoscalar, pseudoscalar potential which does not exist here. The closest we can come to Eq.~(\ref{E1}) is by choosing
the isospin dependent transformation
\begin{equation}
\psi \to \psi' = e^{ibx\gamma_5 \tau_3}\psi .
\label{E3}
\end{equation}
Upon applying this transformation to the stationary version of Eq.~(\ref{B11}) for the vacuum (${\cal D}=1, {\cal C}=0$), we 
generate one chiral spiral with ${\cal D} = e^{2ibx}$ for isospin up and another chiral spiral with ${\cal D}^*=e^{-2ibx}$
for isospin down. These two spirals are winding in opposite directions. The energy values are moved upward or downward by $b$
in the two isospin channels, respectively. The charged pion condensate ${\cal C}$ stays 0, since there is no coupling between the two isospin channels.  
Each isospin component reduces to the standard NJL model. This would yield massless ``baryons" and
a chiral spiral for isospin up and massless ``antibaryons" and the oppositely winding chiral spiral for
isospin down, if these two channels would really be independent. In the isoNJL$_2$ model, the two constructs are locked to each other though, 
connected by the self-consistency condition.
The net result for winding number 1 is a massless, delocalized hadron with $N_c$ ``up-quarks" and $N_c$ ``down-antiquarks"
(or vice versa), carrying zero net baryon number but isospin charge $Q_3=\pm 2N_c$. The spiral case (i.e., a finite density of windings) describes
matter with constant isospin density, but vanishing baryon density. Taken together, the condensates in the two isospin channels may be pictured as a chiral
``double helix". It is impossible to generate fermion number in this manner, due to the absence
of an isoscalar, pseudoscalar interaction term in the Lagrangian (\ref{A2}). The energy density and isospin density can easily be inferred 
from the two separate isospin components to be
\begin{equation}
{\cal E} = \frac{b^2}{2\pi}  N, \quad \rho_3 = \frac{b}{\pi} N, \quad N= 2 N_c.
\label{E4}
\end{equation}
This state is also expected to be thermodynamically favored, establishing a link between thermodynamics of hot and dense
isospin matter ($\mu=0, \mu_3 \neq 0$) in the isoNJL$_2$ model and baryonic matter ($\mu \neq 0$) in the standard NJL$_2$ model.
We therefore expect the phase diagram of the isoNJL$_2$ model in the ($T, \mu_3$) plane (at $\mu=0$)
to be identical to the one of the standard NJL$_2$ model in the ($T, \mu$) plane. 

Let us now comment on the results of the numerical calculation of Heinz et al. \cite{L21}. We have pointed out
in Sect.~\ref{sect4} that the twisted kink which can accommodate the maximum number of fermions in the 
isoNJL$_2$ model is the real GN kink (${\cal D} = \pm \tanh x$ in the rest frame), in both isospin channels.
Dense matter can then most effectively be manufactured as an array of equidistant kinks and antikinks.
This is what happens in the GN model, and the corresponding solution is a self-consistent TDHF solution
of the isoNJL$_2$ model as well. The same arguments go through for the phase diagram as a function of ($T,\mu$).
This is fully consistent with the results of Ref. \cite{L21}. The fact that all 4 matrix elements of $\Delta$
are proportional to the GN kink crystal $S(x)$ then simply reflects the freedom of  performing global chiral rotations,
just like for the vacua. To use other types of crystal with twist different from $\pi$ is disfavored
because the constituent kinks can accommodate fewer fermions. 

What can we say about the phase diagram of the isoNJL$_2$ model with two chemical potentials ($\mu, \mu_3$)?
This has been the subject of Refs. \cite{L18,L19,L20} where various variational ``ans\"atze" have been tried: Homogeneous condensates,
chiral spiral in the scalar-neutral pion sector (${\cal D}$) and homogeneous charged pion condensate (${\cal C}$), chiral spiral in the charged
pion sector and homeogeneous scalar-neutral pion condensate. We should like to propose quite a different picture.
Consider $T=0$ first.
We start from the real kink crystal at $\rho_3=0,\rho \neq 0$ with $\Delta=S, {\cal C}=0$ and $S$ the real
soliton crystal of the GN model with discrete chiral symmetry \cite{L10},
\begin{equation}
S(x) = \kappa \frac{{\rm sn}(x/\kappa){\rm cn}(x/\kappa)}{{\rm dn}(x/\kappa)}.
\label{E5}
\end{equation}
The right hand side is expressed in terms of Jacobi elliptic functions. The elliptic parameter $\kappa$ is related to the density as
\begin{equation}
\rho = \frac{1}{2 \kappa {\bf K}}N .
\label{E6}
\end{equation}
We then apply the above trick by performing a linearly $x$-dependent chiral and isospin-rotation
\begin{equation}
\psi = e^{-ibx\gamma_5 \tau_3} \psi'.
\label{E7}
\end{equation}  
The new spinor $\psi'$ then satisfies the following HF equation,
\begin{equation}
\left( \begin{array}{cccc} i \partial_x & 0 &  {\cal D}^* & 0 \\ 0 & i\partial_x & 0 & {\cal D} \\
{\cal D} & 0 & -i\partial_x & 0 \\ 0 & {\cal D}^* & 0 & -i \partial_x \end{array} \right) \left( \begin{array}{c} \psi_{1,1}' \\ \psi_{1,2}' \\ \psi_{2,1}' \\ \psi_{2,2}'\end{array} \right)
=   \left(  \begin{array}{cccc} \omega + b & 0 & 0 & 0 \\ 0 & \omega-b & 0 & 0 \\ 0 & 0 & \omega+b & 0 \\ 0 & 0 & 0 & \omega-b \end{array} \right)
\left( \begin{array}{c} \psi_{1,1}' \\ \psi_{1,2}' \\ \psi_{2,1}' \\ \psi_{2,2}'\end{array} \right), \quad {\cal D} =S(x)e^{2ibx}.
\label{E8}
\end{equation}
\begin{figure}[h]
\begin{center}
\epsfig{file=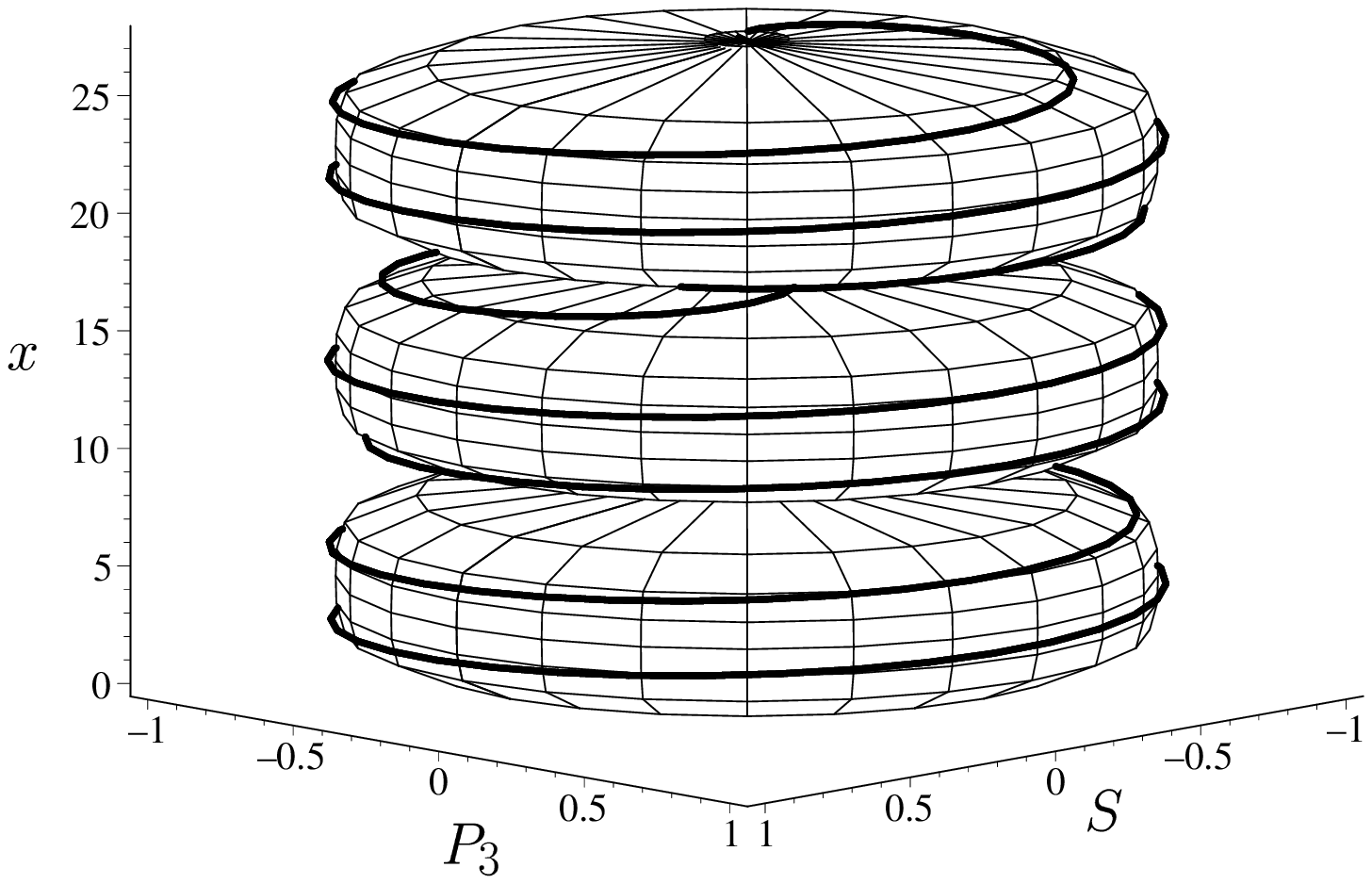,width=6cm,angle=90}
\caption{Example of order parameter ${\cal D}=S-iP_3$ for the isoNJL$_2$ model at finite density and isospin density. The grid is the surface generated by
rotating the real kink crystal order parameter $S(x)$ around the $x$-axis, the fat line is traced out by the order parameter ${\cal D}(x)$.}
\label{fig1}
\end{center}
\end{figure}
The charged pion condensate vanishes, whereas the scalar-neutral pion condensate is the product of the real kink crystal $S$
and the chiral spiral potential $e^{2ibx}$. This yields a distorted double helix, where the radius is no longer constant but also oscillating
in space, see Fig.~\ref{fig1}. Eq.~(\ref{E8}) shows that the isospin up spinor components move up, the isospin down components move down in energy 
by the amount $b$. Evaluating the energy of such a configuration is straightforward, since the effect of the chiral spiral is an ultra violet (UV) effect
independent of the details of the GN soliton crystal,
\begin{eqnarray}
{\cal E}_{\rm isoNJL}(\rho,\rho_3) & = &  {\cal E}_{\rm GN}(\rho) + {\cal E}_{\rm NJL}(\rho_3)   
\nonumber \\
& = & N \left\{ \frac{1}{4\pi} + \frac{1}{\pi \kappa^2} \left( \frac{\bf E}{\bf K} - \frac{1}{2} \right) +\frac{\pi}{2} \left( \frac{\rho_3}{N} \right)^2 \right\}  .
\label{E9}
\end{eqnarray}
At this point we cannot rule out that there exists still a better HF solution, but we find it hard to think of a more economical way of accommodating two different densities of up- and
down-quarks in the isoNJL$_2$ model. Since the chiral spiral trick acts in the UV, we also expect it to be independent of temperature. Our conjecture
for the full phase diagram of the isoNJL$_2$ model is therefore the following: For any given value of $\mu_3$, the phase diagram in the ($\mu, T$) plane
is the same as the one of the GN model with discrete chiral symmetry and baryon chemical potential $\mu$, independently of the value of $\mu_3$, see Fig.~\ref{fig2}.
\begin{figure}[h]
\begin{center}
\epsfig{file=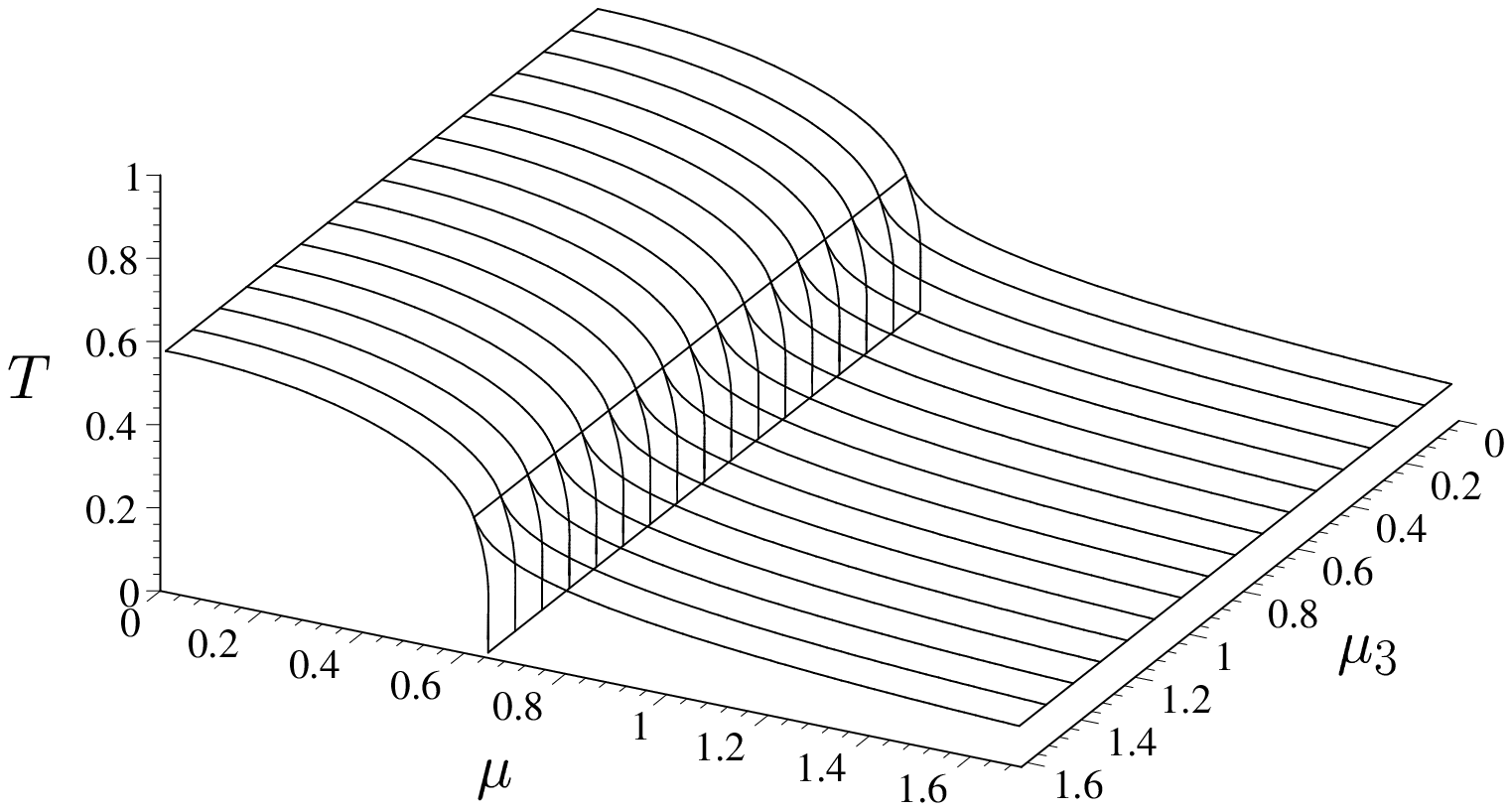,width=6cm,angle=90}
\caption{Phase diagram of the isoNJL$_2$ model as a function of $\mu,\mu_3,T$. In the left hand region under the shaded surface, the radius of the chiral
spiral is constant, the pitch increases with $\mu_3$. In the right hand region, the radius is modulated according to $S(x)$, the pitch also increases
with $\mu_3$. Above the shaded surface, the order parameter vanishes.} 
\label{fig2}
\end{center}
\end{figure}
The charged pion condensate does not
play any role. Of course one could perform a global chiral transformation to the above system and mix in the charged pion condensate, but this would be 
without consequence for the physics.
The mean field at finite temperature should have the same functional form as at $T=0$ but different parameters, which one could presumably infer form
the known phase diagrams of the GN and NJL$_2$ models. This should be a self-consistent HF solution. To prove or disprove our conjecture, or to discover a thermodynamically
more stable solution, it would be interesting to perform a numerical study with temperature and two chemical potentials ($\mu, \mu_3$). 

Finally, let us mention some related work in 3+1 dimensions. In Ref.~\cite{L27}, the NJL model was studied in a magnetic field, at zero temperature.
The authors propose a variational ansatz for the mean field dubbed ``hybrid chiral spiral" which happens to coincide with the form of our ${\cal D}$ in Eq.~(\ref{E8}).
In this case however, the phase rotation is driven by the external field, not the isospin imbalance. Another paper worth mentioning is Ref.~\cite{L28} dealing with
isospin-asymmetric matter in the two-flavor NJL model. The variational ansatz used there is a flavor dependent chiral spiral, with two different pitches for up- and
down quarks. In our case, the periodicity is the same, but the two spirals are winding in opposite directions. The ansatz of Ref.~\cite{L28} also fails to reproduce the
real kink crystal in the isospin symmetric case, which seems to be favored in 3+1 dimensions as well. It may be worthwhile to explore more systematically the
relation between condensates in one and three space dimensions in NJL models, with or without magnetic field.
 
\section{Kink-kink scattering}
\label{sect6}
The most general twisted kink interpolates between two vacua $\Delta_-$ and $\Delta_+$. If $\Delta_-=1$ we will refer to the kink as being ``in the intrinsic
frame".  To transform it to a frame where the initial vacuum is $\Delta_-$ requires only a trivial algebraic operation, hence it is sufficient to
characterize a kink in the intrinsic frame. The SU(2) matrix of the vacuum at $x\to \infty$ in turn can be specified by a ``rotation angle" $\phi$ and 
the direction of a ``rotation axis" $\vec{n}= \vec{\phi}/\phi$. The space-time dependent function $V$ depends on the rotation angle and the kink
velocity, but not on the rotation axis. As we will see, during a kink-kink collision the rotation angle is conserved like in the Abelian case.
The rotation axis will be changed, however.

The single kink problem involves two vacua $\Delta_{\pm}$. We have exploited in Sect.~\ref{sect4} the fact that by a global chiral transformation,
$\Delta_-$ can be mapped to the unit matrix, and $\Delta_+$ can be diagonalized. This enabled us to reduce the problem of the single
non-Abelian twisted kink to the one of the Abelian kink, although with non-trivial implications due to the locking of kink and charge conjugate
kink in the two isospin channels. This trick does not work anymore once we consider the two-kink problem. Here, three distinct vacua are involved before the collision,
namely $\Delta_{\pm}$ at $x\to \pm \infty$ and $\Delta_0$ at $x=0$, say, inbetween the widely separated kinks. During the kink-kink collision, the asymptotic vacua
$\Delta_{\pm}$ remain the same, but $\Delta_0$ will in general change, so that a fourth vacuum comes into play.
We will denote the ``inbetween" vacua $\Delta_0$ before and after the collision as $\Delta_{\rm 0,bef}$ and $\Delta_{\rm 0,aft}$, respectively.
By a global chiral 
transformation we are only able to simplify two out of these four vacua. We shall choose
$\Delta_-$ and $\Delta_{\rm 0,bef}$, transforming $\Delta_-$ to 1 and diagonalizing $\Delta_{\rm 0,bef}$. This is not 
sufficient to decouple the two isospin channels. It is not possible to reduce kink-kink scattering to the corresponding Abelian problem.

Fortunately, a framework has been proposed recently which is perfectly suited to this problem:
It is a general solution of the multi-component BdG equation, derived in the context of exotic 1d superconductors \cite{L25}.
As pointed out by the author, this may be regarded as a generalization of the work of Refs.~\cite{L17,L18} to the case
where the BdG (or, equivalently, Dirac-TDHF) equation involves spinors with several flavor components.
This is exactly what is required here. To adapt this framework to our particular needs, we proceed as follows.
Consider first the single kink in the two-component case as quoted by Takahashi. How does it compare with our twisted kink?
According to Eq.~(2.22) of Ref.~\cite{L25}, this kink interpolates between $\Delta_-=1$ at $x\to -\infty$
and
\begin{equation}
\Delta_+ = 1-2i e^{-i\phi_1}\sin \phi_1 {\cal P}_1
\label{F1}
\end{equation}
at $x\to \infty$, where ${\cal P}_1$ denotes the projector onto the subspace spanned by a complex unit vector $\vec{p}_1$,
\begin{equation}
{\cal P}_1 = \vec{p}_1 \vec{p}_1^{\, \dagger}, \quad \vec{p}_1^{\, \dagger} \vec{p_1} = 1.
\label{F2}
\end{equation}  
$\Delta_+$ is unitary, but not ``special", since det $\Delta_+ = e^{-2i\phi_1}$. Therefore,
the single kink of Ref.~\cite{L25} is not an allowed configuration of the isoNJL$_2$ model
where the vacuum must be an element of SU(2), not U(2).
In order
to see what is missing, let us choose a frame where $\vec{p}_1=(1,0)$. Then the two isospin channels
decouple and 
\begin{equation}
\Delta_+ = \left( \begin{array}{cc} e^{-2i\phi_1} & 0 \\ 0 & 1 \end{array}\right) 
\label{F3}
\end{equation}
What is lacking here is charge conjugation symmetry, enforcing the diagonal elements of $\Delta$
to be complex conjugates $e^{\mp 2i \phi_1}$. If we want nevertheless to use Takahashis
framework, we have to proceed to a bound state of two kinks with twist angle $\phi_1, \phi_2$ and 
complex unit vectors $\vec{p}_1,\vec{p}_2$, but located at the same point in space and moving with the same
velocity. Charge conjugation can then be enforced by requiring the 
vectors $\vec{p}_1,\vec{p}_2$ to be orthogonal and by choosing $\phi_2=\pi-\phi_1$. 
One finds that the resulting potential $\Delta$ interpolates between
$\Delta_-=1$ and 
\begin{equation}
\Delta_+ = e^{-2i\phi_1} {\cal P}_1 + e^{2i \phi_1} {\cal P}_2, \quad {\cal P}_1+{\cal P}_2=1, \quad {\cal P}_1{\cal P}_2=0
\label{F4}
\end{equation}
${\cal P}_1,{\cal P}_2$ are orthogonal projectors, defined via the orthogonal complex vectors $\vec{p}_{1,2}$ as
in Eq.~(\ref{F2}). In the preferred frame, 
\begin{equation}
\vec{p}_1 = \left( \begin{array}{c} 1 \\ 0 \end{array} \right), \quad  \vec{p}_2 = \left( \begin{array}{c} 0 \\ 1 \end{array} \right),
\quad
{\cal P}_1 = \left( \begin{array}{cc} 1 & 0 \\ 0 & 0 \end{array} \right), \quad {\cal P}_2 = \left( \begin{array}{cc} 0 & 0 \\ 0 & 1 \end{array} \right)
\label{F5}
\end{equation}
and hence $\Delta_+ = e^{-2i\phi_1 \tau_3}$. If we rotate the frame as discussed above for the single twisted kink, we see
that the vectors $\vec{p}_1, \vec{p}_2$ can be identified with
\begin{equation}
\vec{p}_1 = \Omega_+^{\dagger} \left( \begin{array}{c} 1 \\ 0 \end{array} \right), \quad  \vec{p}_2 = \Omega_+^{\dagger} \left( \begin{array}{c} 0 \\ 1 \end{array} \right)
\label{F6}
\end{equation} 
with $\Omega_+$ from Eq.~(\ref{D4}).
By comparing the twisted kink of the isoNJL$_2$ model with the (charge conjugation symmetric) two-kink bound state of Takahashi, we now
find perfect agreement. This enables us to solve the kink-kink scattering problem in the isoNJL$_2$ model. 
We need to go to the 4-kink problem (two kink-kink bound states) of the 4-component (2 Dirac, 2 isospin components)
BdG equation. This problem is already fairly complicated. With the help of computer algebra, the results can nevertheless be
reduced to a tractable form. We refer to Ref. \cite{L25} for the details of the calculation to be done.
Moreover, in Appendix 2 we have collected some technical details explaining how to use the formalism of Ref.~\cite{L25} in the present case and establishing the link between
our notation (based on Refs. \cite{L16,L17}) and the one of Takahashi.
Here we proceed directly to some results for kink-kink scattering, adapted to our notation.

To better understand non-Abelian kink-kink scattering, let us first cast the potential in the Abelian case (standard NJL$_2$ model)
of Appendix 1 into the following form,
\begin{equation}
\Delta  =   \frac{1 + e^{-2i\phi_1} V_1 + e^{-2i \phi_2} \xi V_2 + e^{-2i\phi_2} e^{-2i \phi_1} V_1 V_2}{1+V_1+\xi V_2 + V_1V_2}
\label{F7}
\end{equation}
with 
\begin{equation}
V_i =  c_i \exp \left\{ \frac{\sin \phi_i}{\eta_i} \left( \bar{z} + \eta_i^2 z\right) \right\}, \quad \xi = b_{12}^{-1},
\label{F8}
\end{equation}
where $b_{12}$ is given in Eq.~(\ref{Q12}) of Appendix 1.
Consider the asymptotics of incoming and outgoing kinks 1 and 2:
\begin{eqnarray}
\Delta_{\rm 1,in} & = & \lim_{V_2 \to 0} \Delta =  \frac{1+e^{-2i\phi_1}V_1}{1+V_1}
\nonumber \\
\Delta_{\rm 2,in} & = & \lim_{V_1 \to \infty} \Delta = \frac{1 + e^{-2i \phi_2}V_2}{1+V_2} e^{-2i \phi_1}
\nonumber \\
\Delta_{\rm 1,out} & = & \lim_{V_2 \to \infty} \Delta = \frac{1+e^{-2i\phi_1}\xi^{-1} V_1}{1+\xi^{-1} V_1}e^{-2i\phi_2}
\nonumber \\
\Delta_{\rm 2,out} & = & \lim_{V_1 \to 0}\Delta =   \frac{1 + e^{-2i \phi_2} \xi V_2}{1+\xi V_2} 
\label{F9}
\end{eqnarray}
The (real) factor $\xi$ describes the time delay during the collision. 
The following U(1) twist factors (or vacua) can be read off Eq.~(\ref{F9}),
\begin{equation}
\Delta_- = 1, \quad \Delta_+ = e^{-2i\phi_1} e^{-2i\phi_2}, \quad \Delta_{\rm 0,bef} = e^{-2i\phi_1} , \quad \Delta_{\rm 0,aft} = e^{-2i \phi_2}. 
\label{F10}
\end{equation}
Turning to the non-Abelian kinks, we find that the general form of Eq.~(\ref{F7}) is preserved, provided we replace the U(1) twist factors by SU(2) twist matrices,
\begin{equation}
\Delta  =   \frac{1 + U_1 V_1 + U_2 \xi V_2 + U_{12} V_1 V_2}{1+V_1+\xi V_2 + V_1V_2}.
\label{F11}
\end{equation}
Here, $\Delta, U_1,U_2,U_{12}$ are all 2$\times$2 matrices. In the Abelian case, $U_{12}=U_1U_2$, see Eq.~(\ref{F7}). This is not expected to hold in the
non-Abelian case, since the result would depend on the order of the factors. $\xi$ is still a real, scalar factor accounting for
time delay. Its value is different from the Abelian case and will be given below. The space-time dependent functions $V_1,V_2$ are
the same as before, Eq.~(\ref{F8}). This means that the twist angle is
preserved also in the SU(2) case. However the twist axes of both solitons get rotated during the collision. The result 
depends on the initial twist angles, the twist axes and the velocities, and is therefore not a purely geometrical issue.
How the direction of the axes changes during the collision is the main new issue in the non-Abelian case.

In order to simplify the formulas without loss of generality, we go to a Lorentz frame where the kink velocities are equal and
opposite $(\eta_1 = 1/\eta_2 = \eta$). Furthermore we assume that the vacuum at $x\to -\infty$ is $\Delta_-=1$. 
We can diagonalize one further SU(2) matrix, which we choose as $\Delta_{\rm 0,bef}$. The SU(2) matrices $U_1, U_2$ 
can then be represented as
\begin{eqnarray}
U_1 & = & e^{-2i\phi_1 \tau_3},
\nonumber \\
U_2 & = & \Omega^{\dagger} e^{-2i \phi_2 \tau_3} \Omega , \quad \Omega = e^{i \vartheta_2 \tau_2/2}e^{i \varphi_2 \tau_3/2}.
\label{F12}
\end{eqnarray}
The twist angles are $-2\phi_1, -2\phi_2$ and the twist axes
\begin{equation}
\vec{n}_1 = \left( \begin{array}{c} 0 \\ 0 \\ 1 \end{array} \right), \quad \vec{n}_2 = \left( \begin{array}{c} \sin \vartheta_2 \cos \varphi_2 \\
\sin \vartheta_2 \sin \varphi_2 \\ \cos \vartheta_2 \end{array} \right) .
\label{F13}
\end{equation}
In this framework it is easiest to prescribe $U_1,U_2$ (the twists of incoming kink 1 and outgoing kink 2) and compute $U_{12}$, the vacuum at $x\to \infty$.
From the point of view of scattering theory, one would rather prescribe the twists of both incoming kinks, i.e., $U_1,U_{12}$, and compute $U_2$. In order to keep the formulas
as simple as possible, we follow the first road here. Before the collision the sequence of vacua from left to right is $1 \to U_1 \to U_{12}$,
after the collision it is $1 \to U_2 \to U_{12}$, i.e., in the notation of Eq.~(\ref{F10}),
\begin{equation}
\Delta_- = 1, \quad \Delta_+ = U_{12}, \quad \Delta_{\rm 0,bef} = U_1 , \quad \Delta_{\rm 0,aft} = U_2. 
\label{F14}
\end{equation}
The only output is $U_{12}$, from which the intrinsic twists of incoming kink 2 ($U_{12}U_1^{-1}$)
and outgoing kink 1 ($U_{12}U_2^{-1}$) can then be computed. We find the result
\begin{equation}
U_{12} = c_0 + c_1 U_1 + c_2 U_2 + c_{12} U_1U_2 + c_{21} U_2 U_1,
\label{F15}
\end{equation}
as compared to $U_{12} = U_1 U_2$ in the Abelian case. The coefficients are given by
\begin{eqnarray}
c_0 & = & 1- \frac{(1-\eta^4)^2}{\chi \eta^4}
\nonumber \\
c_1 & = & -c_2 = - \frac{2}{\chi} \left( \cos 2 \phi_1 - \cos 2 \phi_2 \right)
\nonumber \\
c_{21} & = & \frac{1-\eta^4}{\eta^4 \chi}
\nonumber \\
c_{12} & = & \frac{\eta^4-1}{\chi}
\nonumber \\
\chi & = & \eta^{-4}+ \eta^4 -2 \left( \cos \vartheta_2 \sin 2 \phi_1 \sin 2 \phi_2 + \cos 2 \phi_1 \cos 2 \phi_2 \right)
\label{F16}
\end{eqnarray}
and  depend on the kink velocity, the twist angles $\phi_1, \phi_2$ and the polar angle $\vartheta_2$ of the rotation axis of $U_2$. The time delay factor $\xi$ now reads
\begin{equation}
\xi = \frac{1}{\eta^4 \chi}\left[ (1+\eta^4)^2-2 \eta^4(\cos 2 \phi_1 + \cos 2 \phi_2)+ 4 \eta^2(1+\eta^4) \sin \phi_1 \sin \phi_2 \right].
\label{F17}
\end{equation}
In the special case of a bound state of two twisted kinks at rest ($\eta = 1$), these expressions simplify to 
\begin{eqnarray}
U_{12} & = &  1 + c_1( U_1 - U_2) 
\nonumber \\
c_1 & = & - \frac{2}{\chi} \left( \cos 2 \phi_1 - \cos 2 \phi_2 \right)
\nonumber \\
\chi & = & 2 \left(1 - \cos \vartheta_2 \sin 2 \phi_1 \sin 2 \phi_2 - \cos 2 \phi_1 \cos 2 \phi_2 \right)
\label{F18}
\end{eqnarray}
Along similar lines one could also investigate more complicated scattering or bound state problems involving more than two kinks,
or breathers with a time dependence in the rest frame. This is outside the scope of the present paper.

\section{Summary and conclusions}
\label{sect7}
We have investigated the variant of the GN model family with non-Abelian chiral symmetry group SU(2)$_L \times$SU(2)$_R$, the isoNJL$_2$ model.
After setting up the TDHF formalism for this model valid in the large $N_c$ limit, we have constructed the non-Abelian twisted kink, interpolating
between two points on the vacuum manifold $S^3$. This novel object may be regarded as a bound state of a standard Abelian kink and its charge 
conjugate, reminiscent of the picture recently developed for the Dashen-Hasslacher-Neveu (DHN) baryon \cite{L29} of the GN model in Ref.~\cite{L17}. 
However, whereas the two constituent kinks of the DHN baryon are separated in space, here they are at the same point in space
but separated in isospin space. Nevertheless many formulas like self-consistency condition, fermion number, bound state energies or 
mass of the non-Abelian twisted kink are strikingly similar to those of the DHN baryon. 

We have then addressed the question of the phase diagram of the isoNJL$_2$ model, focussing in particular on inhomogeneous phases and a
construction analoguous to the chiral spiral in the NJL$_2$ model. Quite a few exact results can be deduced without detailed calculation by
just using known results of the GN and NJL$_2$ models. At zero isospin density or isospin chemical potential $\mu_3$, we could understand
why the phase structure is the same as in the GN model with discrete chiral symmetry, fully confirming recent numerical results from Ref.~\cite{L21}.
For zero baryon density or $\mu=0$, on the other hand, we have given arguments why we expect the phase diagram in the ($T,\mu_3$)-plane to 
coincide with the one of standard
NJL$_2$ model in the ($T,\mu$) plane. It is then not hard to extrapolate these findings to the full ($T,\mu,\mu_3$) space. Our conjecture is
that the mean field in the crystal phase is just the product of the GN and NJL$_2$ mean fields, i.e., a chiral spiral with pitch determined by $\mu_3$
and radius modulated according to the real soliton crystal, depending only on $\mu$. It would be very interesting to check this prediction,
different from all prior variational ``ans\"atze" in the literature on the isoNJL$_2$ model, by means of an unbiased numerical study. 

Finally, we have solved the kink-kink scattering and bound state problem in the NJL$_2$ model, using a formalism recently developed in
condensed matter physics for the multi-component BdG equation \cite{L25}. We find that the twist angle is conserved in the collision, but 
the twist axes are rotated in a complicated way. This is where the non-Abelian character of the kinks shows up most clearly. 
One could now go on and study breathers or more complicated bound state and scattering problems, but this is beyond the scope
of the present, exploratory study. In any case, there is no doubt that the isoNJL$_2$ model is also integrable. The formalism
of Ref.~\cite{L25}) could even be used to generalize this study to higher chiral groups like SU(3)$_L \times$SU(3)$_R$.

\section*{Appendix 1: Abelian kinks in the NJL$_2$ model}

Here we collect the main results for one and two twisted kinks of the standard NJL$_2$ model, obtained using Refs.~\cite{L16,L17}. 
Light cone coordinates,
\begin{equation}
z=x-t, \quad \bar{z} =x+t, \quad \partial_0 = \bar{\partial}-\partial, \quad \partial_1 = \bar{\partial}+\partial.
\label{Q1}
\end{equation}
Single particle spectral parameter $\zeta$, momentum $k$ and energy $E$,
\begin{equation}
k = \frac{1}{2} \left( \zeta- \frac{1}{\zeta} \right), \quad E = - \frac{1}{2} \left( \zeta + \frac{1}{\zeta} \right).
\label{Q2}
\end{equation}
Covariant form of Dirac-TDHF equation
\begin{equation}
2i \bar{\partial} \psi_2 = \Delta \psi_1, \quad 2i \partial \psi_1 = - \Delta^* \psi_2, \quad \Delta=S-iP.
\label{Q3}
\end{equation}
Continuum spinor,
\begin{equation}
\psi_{\zeta} = \frac{1}{\sqrt{1+\zeta^2}} \left( \begin{array}{c} \zeta \chi_1 \\ - \chi_2 \end{array} \right)
e^{i(\zeta \bar{z}-z/\zeta)/2}.
\label{Q4}
\end{equation}
Mean field $\Delta$ for one twisted kink,
\begin{eqnarray}
\Delta & = &  \frac{1+ \frac{\zeta_1}{\zeta_1^*} V_1}{1+V_1} =  \frac{1+ e^{-2i \phi_1} V_1}{1+V_1} ,
\nonumber \\
V_1 & = &  c_1 \exp \left\{ \frac{\sin \phi_1}{\eta_1} \left( \bar{z} + \eta_1^2 z\right) \right\}
\nonumber \\
& = & c_1 \exp \left\{  2 \sin \phi_1 \frac{x-v_1t}{\sqrt{1-v_1^2}} \right\} ,
\label{Q5}
\end{eqnarray}
with $c_1$ a real constant and 
\begin{eqnarray}
\zeta_1 & = &  - \frac{e^{-i \phi_1}}{\eta_1},
\nonumber \\
\eta_1 & = &  e^{\xi_1} = \sqrt{ \frac{1+v_1}{1-v_1}},
\label{Q6}
\end{eqnarray}
(rapidity $\xi_1$, velocity $v_1$). We ususally refer to $\phi_1$ as twist angle, although the twist is actually $-2\phi_1$, see Eq.~(\ref{Q5}).
Continuum spinor for one kink: Eq.~(\ref{Q4}) with
\begin{equation}
\chi_1  =  \frac{1 + \frac{\zeta-\zeta_1^*}{\zeta- \zeta_1} V_1}{1+V_1},
\quad
\chi_2  =  \frac{1+ \frac{\zeta_1}{\zeta_1^*}\frac{\zeta-\zeta_1^*}{\zeta-\zeta_1} V_1}{1+V_1}.
\label{Q7}
\end{equation}
Normalized bound state spinor
\begin{eqnarray}
\varphi_1 & = & \frac{1}{\sqrt{2 \sin \phi_1}} \frac{e_1}{1+V_1}, \quad \varphi_2 = - \frac{1}{\zeta_1^*} \varphi_1,
\nonumber \\
e_1 & = &  e^{i(\zeta_1^* \bar{z}-z/\zeta_1^*)/2} .
\label{Q8}
\end{eqnarray}
Bound state energy in the rest frame,
\begin{equation}
E_0 = \cos \phi_1.
\label{Q9}
\end{equation}
Occupation fraction of the bound state (equivalent to self-consistency condition),
\begin{equation}
\nu_1 = \frac{\phi_1}{\pi}.
\label{Q10}
\end{equation}
The fermion density vanishes identically.

For the charge conjugate solution, make the following replacements,
\begin{eqnarray}
\Delta & \to & \Delta^*,
\nonumber \\
\psi_{\zeta} & \to & \gamma_5 \psi_{-\zeta}^*,
\nonumber \\
\varphi_1 & \to  & \varphi_1^*, \quad \varphi_2 \to - \varphi_2^*. 
\label{Q11}
\end{eqnarray}
The twist angle $\phi_1$ has to be replaced by $\pi-\phi_1$, $\zeta_1$ by $-\zeta_1^*$.

Results for kink-kink scattering: TDHF potential,
\begin{eqnarray}
\Delta & = & D^{-1}\left(1+ e^{-2i\phi_1} V_1 + e^{-2i\phi_2} V_2 + b_{12} e^{-2i(\phi_1+\phi_2)}V_1V_2 \right)
\nonumber \\
& = & D^{-1} \left( 1+ \frac{\zeta_1}{\zeta_1^*}V_1 + \frac{\zeta_2}{\zeta_2^*} V_2 + b_{12} \frac{\zeta_1}{\zeta_1^*}\frac{\zeta_2}{\zeta_2^*}   V_1V_2\right)
\nonumber \\
D & = & 1+V_1+V_2+b_{12} V_1V_2
\nonumber \\
b_{12} & = &  \left| \frac{\zeta_1-\zeta_2}{\zeta_1-\zeta_2^*} \right|^2 = \frac{\eta_1^2+\eta_2^2-2 \eta_1 \eta_2 \cos (\phi_1-\phi_2)}
{\eta_1^2+\eta_2^2-2 \eta_1 \eta_2 \cos (\phi_1+\phi_2)}
\label{Q12}
\end{eqnarray}
Scattering states,
\begin{eqnarray}
\chi_1 & = & D^{-1} \left( 1+ \frac{\zeta-\zeta_1^*}{\zeta-\zeta_1} V_1 +  \frac{\zeta-\zeta_2^*}{\zeta-\zeta_2}  V_2 + b_{12} 
\frac{\zeta-\zeta_1^*}{\zeta-\zeta_1}  \frac{\zeta-\zeta_2^*}{\zeta-\zeta_2}  V_1 V_2\right) 
\nonumber \\
\chi_2 & = & D^{-1} \frac{\zeta_1}{\zeta_1^*} \left(1+ \frac{\zeta-\zeta_1^*}{\zeta-\zeta_1} V_1 +  \frac{\zeta_2}{\zeta_2^*} \frac{\zeta-\zeta_2^*}{\zeta-\zeta_2}  V_2 + b_{12} 
 \frac{\zeta_1}{\zeta_1^*}  \frac{\zeta_2}{\zeta_2^*} \frac{\zeta-\zeta_1^*}{\zeta-\zeta_1}  \frac{\zeta-\zeta_2^*}{\zeta-\zeta_2}  V_1 V_2 \right) 
\label{Q13}
\end{eqnarray}
Bound states,
\begin{eqnarray}
\varphi^{(1)} & = & \frac{1}{D\sqrt{2 \sin \phi_1}} \left( \begin{array}{c} e_1 \left( 1 + \frac{\zeta_1^*-\zeta_2^*}{\zeta_1^*- \zeta_2} V_2 \right) \\
- \frac{e_1}{\zeta_1^*} \left( 1 + \frac{\zeta_2}{\zeta_2^*} \frac{\zeta_1^*-\zeta_2^*}{\zeta_1^*- \zeta_2} V_2 \right) \end{array} \right) 
\nonumber \\
\varphi^{(2)} & = & \frac{1}{D\sqrt{2 \sin \phi_2}} \left( \begin{array}{c} e_2 \left( 1 + \frac{\zeta_1^*-\zeta_2^*}{\zeta_1- \zeta_2^*} V_1 \right) \\
- \frac{e_2}{\zeta_2^*} \left( 1 + \frac{\zeta_1}{\zeta_1^*} \frac{\zeta_1^*-\zeta_2^*}{\zeta_1- \zeta_2^*} V_1 \right) \end{array} \right) 
\label{Q14}
\end{eqnarray}
The denominators $D$ in (\ref{Q13},\ref{Q14}) are the same as in (\ref{Q12}).

\section*{Appendix 2: Solving the isoNJL$_2$ model with the method of Ref.~\cite{L25}}

Here we briefly describe how to use the formalism developed for the multicomponent BdG equation by Takahashi \cite{L25} for solving the
one- and two-kink problem of the isoNJL$_2$ model. Actually, essentially all what is needed from Ref.~\cite{L25} is contained in Sect. IIA. 

{\em One kink problem:}

To get the simple, diagonal form (\ref{D8}) of the non-Abelian twisted kink, set up the 4$\times$2 matrix
\begin{equation}
W(x,t) = \left( \begin{array}{cc} e_1  & 0 \\ 0 & e^*_1 \\ - \frac{e_1}{\zeta_1^*} & 0 \\ 0 & \frac{e_1^*}{\zeta_1}\end{array} \right)
\label{R1}
\end{equation}  
with 
\begin{equation}
\zeta_1 = - \frac{e^{-i\phi_1}}{\eta_1}, \quad e_1 = e^{i(\zeta_1^* \bar{z}-z/\zeta_1^*)/2}.
\label{R2}
\end{equation} 
Light cone coordinates are defined in Eq.~(\ref{Q1}) of Appendix 1. Construct a 2$\times$2 matrix $G$ and a 4$\times$2 matrix $H$ as follows,
\begin{eqnarray}
G(x,t) & = & \int_{- \infty}^x dy W^{\dagger}(y,t) W(y,t),
\nonumber \\
H(x,t) & = & -W(x,t) \left( 1 + G(x,t) \right)^{-1},
\label{R3}
\end{eqnarray}
as well as the 4$\times$4 matrix
\begin{equation}
K(x,t)=H(x,t)W^{\dagger}(x,t).
\label{R4}
\end{equation} 
Introduce the 4$\times$4 matrices
\begin{equation}
A= \left( \begin{array}{cccc} 1 & 0 & 0 & 0 \\ 0 & 1 & 0 & 0 \\ 0 & 0 & -1 & 0 \\ 0 & 0 & 0 & -1 \end{array} \right), \quad
B = \left( \begin{array}{cccc} 0 & 0 & -i & 0 \\ 0 & 0 & 0 & -i \\ -i & 0 & 0 & 0 \\ 0 & -i & 0 & 0 \end{array} \right).
\label{R5}
\end{equation}
Then
\begin{equation}
i \left(B + [ K,A ]\right) = \left( \begin{array}{cc} 0 & \Delta^{\dagger} \\ \Delta & 0 \end{array} \right)
\label{R6}
\end{equation}
The left hand side is a 4$\times$4 matrix constructed according to the rules taken from \cite{L25} as summarized above. The right hand side
is a 4$\times$4 matrix written in 2$\times$2 block form. The lower left 2$\times$2 block labelled $\Delta$ gives the TDHF potential for
the non-Abelian kink of the isoNJL$_2$ model. In order to reproduce the form given in the main text,
one has to define
\begin{equation}
V_1= \frac{\eta_1}{\sin \phi_1} |e_1|^2.
\label{R7}
\end{equation}
The normalized spinors for the two bound states are given by the two columns of the matrix $H$. Continuum spinors
in our standard normalization can finally be obtained as follows: Define yet another 4$\times$2 matrix
\begin{equation}
\Phi(x,t) =  \left( \begin{array}{cc} \zeta & 0 \\ 0 & \zeta \\ -1 & 0 \\ 0 & -1 \end{array} \right) \frac{e^{i(\zeta \bar{z}-z/\zeta)/2}}{\sqrt{1+\zeta^2}} .
\label{R8}
\end{equation}
The continuum spinors in the two isospin channels are then the two columns of the 4$\times$2 matrix
\begin{equation}
\Psi(x,t) = \Phi(x,t) + H(x,t) \int_{- \infty}^x dy W^{\dagger}(y,t) \Phi(y,t).
\label{R9}
\end{equation}

{\em Two kink problem:}

To derive the TDHF potential for kink-kink scattering in the frame used in Sect.~\ref{sect6} of the main text, set up the 4$\times$4 matrix
\begin{equation}
W(x,t) = \left( \begin{array}{cccc} e_1  & 0 & e_2 p_{11} & e_2^* p_{21} \\ 0 & e_1^* & e_2 p_{12} & e_2^* p_{22}  \\ - \frac{e_1}{\zeta_1^*} & 0 &  -\frac{e_2}{\zeta_2^*} p_{11} & 
\frac{e_2^*}{\zeta_2} p_{21} \\ 0 & \frac{e_1^*}{\zeta_1} &  - \frac{e_2}{\zeta_2^*} p_{12} & 
\frac{e_2^*}{\zeta_2} p_{22}  \end{array} \right).
\label{R10}
\end{equation}  
Two complex, orthogonal, normalized vectors $\vec{p}_1, \vec{p}_2$ have been introduced as follows,
\begin{eqnarray}
\vec{p}_1 & = &  \left( \begin{array}{c} p_{11} \\ p_{12} \end{array} \right) = \Omega^{\dagger} \left( \begin{array}{c} 1 \\ 0 \end{array} \right), 
\nonumber \\
\vec{p}_2 & = &  \left( \begin{array}{c} p_{21} \\ p_{22} \end{array} \right) = \Omega^{\dagger} \left( \begin{array}{c} 0 \\ 1 \end{array} \right), 
\nonumber \\
\Omega & = & e^{i \vartheta_2 \tau_2/2} e^{i \varphi_2 \tau_3/2} 
\label{R11}
\end{eqnarray}
We use
\begin{equation}
\zeta_1 = - \frac{e^{-i\phi_1}}{\eta}, \quad \zeta_2 = - \eta e^{-i\phi_2}, \quad e_i = e^{i(\zeta_i^* \bar{z}-z/\zeta_i^*)/2}.
\label{R12}
\end{equation}
$G,H$ and $K$ are then defined as in Eqs.~(\ref{R3},\ref{R4}), but now all become 4$\times$4 matrices. 
Eqs.~(\ref{R5},\ref{R6}) also remain valid whereas (\ref{R7}) has to be replaced by 
\begin{equation}
V_1 = \frac{\eta}{\sin \phi_1} |e_1|^2, \quad \xi V_2 = \frac{1}{\eta \sin \phi_2} |e_2|^2.
\label{R13}
\end{equation}
The only technical difficulty is the fact that by applying this
formalism, one generates the square of the denominator of Eq.~(\ref{F11}), and a correspondingly more complicated
numerator. Since these expression are very lengthy and not in factorized form, most of the computer algebra effort actually 
goes into the factorization to arrive at the simple expression (\ref{F11}). 

There are now 4 bound states. Their
normalized spinors can be read off the 4 columns of the matrix $H$. The continuum spinors are constructed as 
in Eqs. (\ref{R7},\ref{R8}). Both bound state and continuum spinors suffer from the same disease as the TDHF potential,
i.e., they need to be factorized. Since we have not 
found any compact way of writing down the final, factorized expressions for the spinors, we do not show them
in the present paper, but give only the TDHF potential $\Delta$.



\begin{thebibliography}{99}
\bibitem{L1}
D. J. Gross and A. Neveu, Phys. Rev. D {\bf 10}, 3235 (1974).
\bibitem{L2}
Y. Nambu and G. Jona-Lasinio, Phys. Rev. {\bf 124}, 246 (1961).
\bibitem{L3}
L. McLerran and R. D. Pisarski, Nucl. Phys. A {\bf 796}, 83 (2007).
\bibitem{L4}
T. Kojo, Y. Hidaka, L. McLerran, R. D. Pisarski, Nucl. Phys. A {\bf 843}, 37 (2010).
\bibitem{L5}
T. Kojo, Y. Hidaka, K. Fukushima, L. D. McLerran, R. D. Pisarski, Nucl. Phys. A {\bf 875}, 94 (2012).
\bibitem{L6}
V. Sch\"on and M. Thies, Phys. Rev. D {\bf 62}, 096002 (2000).
\bibitem{L7}
D. Nickel, Phys. Rev. D {\bf 80}, 074025 (2009).
\bibitem{L8}
S. Carignano, D. Nickel, M. Buballa, Phys. Rev. D {\bf 82}, 054009 (2010).
\bibitem{L9}
M. Buballa, S. Carignano, Prog. Part. Nucl. Phys. {\bf 81}, 39 (2015).
\bibitem{L10}
M. Thies, Phys. Rev. D {\bf 69}, 067703 (2004).
\bibitem{L11}
M. Thies, J. Phys. A: Math. Gen. {\bf 39}, 12707 (2006).
\bibitem{L12}
V. Sch\"on and M. Thies, {\em At the frontiers of particle physics: Handbook of QCD, Boris Ioffe Festschrift}, edited by M. Shifman
(World Scientific, Singapore, 2001), Vol. 3, p. 1945.
\bibitem{L13}
G. Basar, G. V. Dunne, M. Thies, Phys. Rev. D {\bf 79}, 105012 (2009).
\bibitem{L14}
D. A. Takahashi and M. Nitta, Phys. Rev. Lett. {\bf 110}, 131601 (2013).
\bibitem{L15}
G. V. Dunne and M. Thies, Phys. Rev. Lett. {\bf 111}, 121602 (2013).
\bibitem{L16}
G. V. Dunne and M. Thies, Phys. Rev. A {\bf 88}, 062115 (2013).
\bibitem{L17}
G. V. Dunne and M. Thies, Phys. Rev. D {\bf 89}, 025008 (2014).
\bibitem{L18}
D. Ebert and K. G. Klimenko, {\em Pion condensation in the Gross-Neveu model with nonzero baryon and isospin chemical potentials},
arXiv:0902.1861 [hep-ph].
\bibitem{L19}
D. Ebert, N. V. Gubina, K. G. Klimenko, S. G. Kurbanov, V. Ch. Zhukovsky, Phys. Rev. D {\bf 84}, 025004 (2011).
\bibitem{L20}
N. V. Gubina, K. G. Klimenko, S. G. Kurbanov, V. Ch. Zhukovsky, Phys. Rev. D {\bf 86}, 085011 (2012).
\bibitem{L21}
A. Heinz, F. Giacosa, M. Wagner, D. H. Rischke, {\em Inhomogeneous condensation in effective models for QCD
using the finite-mode approach}, arXiv:1508.06057 [hep-ph].
\bibitem{L22}
M. Thies and K. Urlichs, Phys. Rev. D {\bf 67}, 125015 (2003).
\bibitem{L23}
O. Schnetz, M. Thies, K. Urlichs, Ann. Phys. {\bf 314}, 425 (2004).
\bibitem{L24}
S.-S. Shei, Phys. Rev. D {\bf 14}, 535 (1976).
\bibitem{L25}
D. A. Takahashi, Phys. Rev. B {\bf 93}, 024512 (2016). 
\bibitem{L26}
F. Karbstein and M. Thies, Phys. Rev. D {\bf 76}, 085009 (2007).
\bibitem{L27}
K. Nishiyama, S. Karasawa, T. Tatsumi, Phys. Rev. D {\bf 92}, 036008 (2015).
\bibitem{L28}
D. Nowakowski, M. Buballa, S. Carignano, J. Wambach, {\em Inhomogeneous chiral symmetry breaking phases in isospin-asymmetric
matter}, arXiv:1506.04260 [hep-ph].
\bibitem{L29}
R. F. Dashen, B. Hasslacher, A. Neveu, Phys. Rev. D {\bf 12}, 2443 (1975).
\end{thebibliography}
\end{document}